\documentclass[final,3p,times,twocolumn]{elsarticle}

\usepackage{paralist, enumerate, enumitem}
\usepackage{amssymb, amsmath, amsthm, bm}
\usepackage{geometry}
\usepackage{graphicx, subfigure}
\usepackage{booktabs, multirow, tabu}
\usepackage[switch, mathlines]{lineno}
\newlength\imagewidth
\newtheorem{Property}{Property}

\journal{Optics $\&$ Laser Technology}
\newcommand\Comp{{\rm F}}
\newcommand\Map{{\rm Map}}

\begin{document}

\begin{frontmatter}

\title{Cryptanalyzing a RGB image encryption algorithm based on DNA encoding and chaos map}

\author[cn-xtu-cie]{Yuansheng Liu\corref{corr}}
\ead{yyuanshengliu@gmail.com}

\cortext[corr]{Corresponding author.}
\address[cn-xtu-cie]{College of Information Engineering, Xiangtan University, Xiangtan 411105, Hunan, China}

\begin{abstract}
Recently, a RGB image encryption algorithm based on DNA encoding and chaos
map has been proposed. It was reported that the encryption algorithm can be broken
with four pairs of chosen plain-images and the corresponding cipher-images.
This paper re-evaluates the security of the encryption algorithm, and finds
that the encryption algorithm can be broken efficiently with only
one known plain-image.
The effectiveness of the proposed known-plaintext attack
is supported by both rigorous theoretical analysis and experimental results. In addition, two other security defects
are also reported.
\end{abstract}

\begin{keyword}
image encryption\sep cryptanalysis \sep known-plaintext attack
\end{keyword}

\end{frontmatter}

\linenumbers

\section{Introduction}
\label{sec:intro}

With the development of communication and social networking
technologies, especially the popularity of smart phones,
image transmission over network occur more and more frequently.
Correspondingly, the security of digital images becomes more and
more important. The traditional text encryption techniques cannot
protect images efficiently because of the fundamental characteristics
of images, such as bulk data capacity and high correlation among
pixels. The intrinsic features of chaos, such as ergodicity, mixing
property, sensitivity to initial conditions and control parameters
\cite{Alvarez2006some}, attract researchers to consider chaos
as a novel way to design secure and efficient encryption algorithms
\cite{fridrich1998symmetric, chen2004symmetric, masuda2006chaotic,
chen2011modified, zhu2012novel, jakimoski2001chaos}.
Meanwhile, some cryptanalysis work \cite{li2012breaking,
li2011breaking, li2009security, solak2010cryptanalysis, li2008cryptanalysis,
rhouma2010cryptanalysis}
have been found many of them have security problems from the modern cryptographical point view.

Due to the vast parallelism and extraordinary information density
exclusive characteristic of DNA molecule, and
the rapid development of DNA computing \cite{DNAComputingAndApp},
DNA cryptography \cite{xiao2006new, zhang2012research} has infiltrated
into the field of cryptography. A number of image encryption algorithms
with the idea of combining chaos and DNA computing have been developed
\cite{awad2012new, soni2012encryption, zhang2013couple,
wei2012novel, zhang2010novel, zhang2013novel, zhang2010image}.
The kernel of these algorithms is DNA encoding and DNA computing
which include some biological operations and algebra operations
on DNA sequence, such as the complementary rule of bases
\cite{zhang2010image, Liu20121457}, DNA addition
\cite{zhang2013couple, wei2012novel, soni2012encryption},
DNA exclusive OR operation \cite{zhang2010novel, zhang2013novel}.
Recent cryptanalysis results \cite{hermassi2013security} have shown
that the algorithm proposed in \cite{zhang2010image}
is non-invertible and insecure against chosen-plaintext attack.
Zhang et al. \cite{Zhang2013} found that the encryption algorithm proposed
in \cite{zhang2013novel} can be broken by choosing $(4mn/3 + 1)$ pairs of
plain-images, where $mn$ is the size of the plain-image.

In \cite{Liu20121240}, a RGB image encryption algorithm
based on DNA encoding and chaos map was proposed.
Shortly after the publication of the encryption algorithm,
{\"O}zkaynak et al. \cite{ozkaynakkaos} found that
the encryption algorithm is insecure against chosen-plaintext
attack and the equivalent secret key of the encryption algorithm
can be obtained by four chosen plain-images. This
paper re-evaluates the security of the image encryption
algorithm proposed in \cite{Liu20121240}, and discovers
the following security problems:
(1) the equivalent secret key of the encryption algorithm can
be reconstructed with only one pair of known-plaintext/ciphertext;
(2) encryption results are not sensitive with respect to changes of the
plain-images/secret key.

The remaining of the paper is organized as follows. In the next section,
we will give a detailed introduction of the image encryption algorithm under study.
Section~\ref{sec:ca} describes the proposed known-plaintext attack in detail
with some experimental results and reports the other two security defects.
The last section concludes the paper.

\section{The image encryption algorithm under study}
\label{sec:alogrithm}

\begin{table*}[htb]
\centering
\caption{Eight DNA map rules.}
\begin{tabular}{ccccccccc}
\toprule &$1$   &$2$   &$3$   &$4$   &$5$   &$6$   &$7$   &$8$ \\
\midrule &$0-A$ &$0-A$ &$0-C$ &$0-C$ &$0-G$ &$0-G$ &$0-T$ &$0-T$\\
         &$1-C$ &$1-G$ &$1-A$ &$1-T$ &$1-A$ &$1-T$ &$1-C$ &$1-G$\\
         &$2-G$ &$2-C$ &$2-T$ &$2-A$ &$2-T$ &$2-A$ &$2-G$ &$2-C$\\
         &$3-T$ &$3-T$ &$3-G$ &$3-G$ &$3-C$ &$3-C$ &$3-A$ &$3-A$\\
\bottomrule
\end{tabular}
\label{tab:EncodingRules}
\end{table*}

\begin{table}[htb]
\centering
\caption{DNA addition and subtraction operation.}
\begin{tabular}{ccccccccccc}
\toprule $+$  &$A$  &$T$  &$C$  &$G$ &$$ &$-$  &$A$  &$T$  &$C$  &$G$\\
\midrule $A$  &$T$  &$G$  &$A$  &$C$ &$$ &$A$  &$C$  &$G$  &$A$  &$T$\\
         $T$  &$G$  &$C$  &$T$  &$A$ &$$ &$T$  &$A$  &$C$  &$T$  &$G$\\
         $C$  &$A$  &$T$  &$C$  &$G$ &$$ &$C$  &$G$  &$T$  &$C$  &$A$\\
         $G$  &$C$  &$A$  &$G$  &$T$ &$$ &$G$  &$T$  &$A$  &$G$  &$C$\\
\bottomrule
\end{tabular}
\label{tab:AdditionAndSubtraction}
\end{table}

The plaintext of the encryption algorithm under study is a
RGB color image of size $H\times W$ (height$\times$width),
which is scanned in the raster order and represented as
a one dimensional sequence 
$\mathbf{I}= \{I_i\}_{i=1}^{L}=\{(R_i, G_i, B_i)\}_{i=1}^{L}$,
where $L=H\times W$.
Then, a sequence 
$\mathbf{I}_b = \{(r_i, g_i, b_i)\}_{i=1}^{4L}$ is constructed,
where $(\sum_{k=0}^{3}r_{4\cdot i-k}\cdot 4^{k},
\sum_{k=0}^{3}g_{4\cdot i-k}\cdot 4^{k}, \sum_{k=0}^{3}b_{4\cdot i-k}\cdot 2^{k})
=(R_i, G_i, B_i)$.
The cipher operates on $\mathbf{I}_b$ and gets
$\mathbf{I}'_b = \{(r'_i, g'_i, b'_i)\}_{i=1}^{4L}$.
Finally, the cipher-image $\mathbf{I}'=\{I'_i\}_{i=1}^{L}=\{(R'_i, G'_i, B'_i)\}_{i=1}^{L}$
is obtained via $(R'_i, G'_i, B'_i)=
(\sum_{k=0}^{3}r'_{4\cdot i-k}\cdot 4^{k}, \sum_{k=0}^{3}g'_{4\cdot i-k}\cdot 4^{k},
\sum_{k=0}^{3}b'_{4\cdot i-k}\cdot 4^{k})$.
In the encryption algorithm DNA coding rule performed as a part of secret key and
DNA addition operation is used to confuse the DNA sequences.
Eight DNA map rules that satisfy the Watson-Crick complement rule and
the detail of addition and subtraction operations are shown in
Table~\ref{tab:EncodingRules} and Table~\ref{tab:AdditionAndSubtraction},
respectively. With these preliminary introduction, the image encryption
algorithm under study can be described in detail as follows\footnote{
To simplify the description of the encryption algorithm
under study, some notations in the original paper \cite{Liu20121240}
are modified under the condition that the encryption algorithm is not changed.}.

\begin{itemize}
\item \textit{The secret key} is composed of two DNA map rules $k_1, k_2\in [1, 8]$,
and two sets of initial condition and control parameter of the logistic map
\begin{linenomath}
\begin{equation}
x_{i+1}=\mu \cdot x_i \cdot (1-x_i),
\label{eq:logistic}
\end{equation}
\end{linenomath}
$(x_0, \mu_0), (x'_0, \mu'_0)$, where $x_0, x'_0\in (0, 1)$, and
$\mu_0, \mu'_0\in(3.569945, 4)$.

\item \textit{The initialization procedure}

(1) Iterate the logistic map~(\ref{eq:logistic}) $4L$ times to obtain
a chaotic states sequence $\{S_i\}_{i=1}^{4L}$ with the set of initial condition
and control parameter $(x_0, \mu_0)$.
For $i=1\sim 4L$, set
\begin{linenomath}
\begin{equation}
\nonumber
z_i =
\begin{cases}
0, & \mbox{if } 0 < S_i \leq 0.5,\\
1, & \mbox{if } 0.5 < S_i < 1.\\
\end{cases}
\end{equation}
\end{linenomath}

(2) Iterate the logistic map~(\ref{eq:logistic}) $L$ times to obtain
a chaotic states sequence $\{S'_i\}_{i=1}^{L}$ with the set of initial condition
and control parameter $(x'_0, \mu'_0)$.
For $i=1\sim L$, set
\begin{linenomath}
\begin{equation}
\nonumber
T_i=(\lfloor S'_i\times 10^{5}\rfloor) \mod 256,
\end{equation}
\end{linenomath}
where $\lfloor x\rfloor$ round $x$ to the nearest integers less than or
equal to $x$. Then,
a sequence $\{t_i\}_{i=1}^{4L}$ can be constructed,
where $\sum_{k=0}^{3}t_{4\cdot i-k}\cdot 4^{k}=T_i$.

\item \textit{The encryption procedure} consists of the following five
steps.
\begin{itemize}
\item \textit{Step (a) Encoding}. The DNA map rule $k_1$ is employed
to encode $\mathbf{I}_b$ and then get a DNA sequence
$\{(D^r_i, D^g_i, D^b_i)\}_{i=1}^{4L}$.

\item \textit{Step (b) Addition}. For $i=1\sim 4L$, set
\begin{linenomath}
\begin{equation}
\nonumber
(N^r_i, N^g_i, N^b_i) = (D^r_i + D^g_i, D^g_i + D^b_i, N^g_i + D^b_i).
\end{equation}
\end{linenomath}

\item \textit{Step (c) Complement}. For $i=1\sim 4L$, set
\begin{linenomath}
\begin{equation}
\nonumber
(R^r_i, R^g_i, R^b_i) =
\begin{cases}
(N^r_i, N^g_i, N^b_i), &\mbox{if } z_i=0,\\
(\Comp(N^r_i), \Comp(N^g_i), \Comp(N^b_i)), &\mbox{if } z_i=1,\\
\end{cases}
\end{equation}
\end{linenomath}
where
\begin{linenomath}
\begin{equation}
\label{eq:comp}
\Comp(X) =
\begin{cases}
T, &\mbox{if } X=A,\\
C, &\mbox{if } X=G,\\
G, &\mbox{if } X=C,\\
A, &\mbox{if } X=T.\\
\end{cases}
\end{equation}
\end{linenomath}

\item \textit{Step (d) Decoding}. Decode $\{(R^r_i, R^g_i, R^b_i)\}_{i=1}^{4L}$ to
get a sequence
$\mathbf{I}^*= \{(r^*_i, g^*_i, b^*_i)\}_{i=1}^{4L}$ with the DNA map rule $k_2$.

\item \textit{Step (e) Masking}. For $i=1\sim 4L$, set
\begin{linenomath}
\begin{equation}
\label{eq:masking}
(r'_i, g'_i, b'_i) = (r^*_i \oplus t_i, g^*_i \oplus t_i, b^*_i \oplus t_i),
\end{equation}
\end{linenomath}
where $\oplus$ denotes the bitwise exclusive OR operation.
\end{itemize}

\item \textit{The decryption procedure} is the simple reversion of
the above encryption procedure.

\end{itemize}

\section{Cryptanalysis}
\label{sec:ca}

\subsection{Known-plaintext attack}
\label{subsec:kpa}

The known-plaintext attack is a cryptanalysis model which the
attacker has some samples of both the plaintext and the
corresponding ciphertext. The goal of the attack is to
reveal some secret information, such as secret keys and/or its
equivalent ones. Strength of the encryption algorithm against the
known-plaintext attack is one of the most important factors to
evaluate its security. Under the known-plaintext attack,
the image algorithm under study can be broken with only one
plain-image and its corresponding cipher-image.

Before introducing the known-plaintext attack, two properties
of the encryption algorithm are given, which are the core of the
proposed attack.

\begin{Property}
\label{pro:bijective}
The encryption procedures of \textit{Step (c)} to \textit{Step (e)} are equivalent
to the following operation:
\begin{linenomath}
\begin{equation}
\nonumber
(r'_i, g'_i, b'_i) = (f_i(N^r_i), f_i(N^g_i), f_i(N^b_i)),
\end{equation}
\end{linenomath}
where $i\in [1, 4L]$ and $f_i$ is a bijective
$f_i: \{A, G, C, T\}\to \{0, 1, 2, 3\}$,
and $f_i$ is equivalent to a DNA map rule.
\end{Property}
\begin{proof}
\begin{inparaenum}[(1)]
\item From Eq.~(\ref{eq:comp}), one can know that the operation of
\textit{Step (c) Complement} is a bijective
$g_i^1: \{A, G, C, T\}\to \{A, G, C, T\}$.
\item From table~\ref{tab:EncodingRules}, one can consider the operation
of \textit{Step (d) Decoding} as a bijective
$g_i^2: \{A, G, C, T\}\to \{0, 1, 2, 3\}$.
\item From Eq.~(\ref{eq:masking}), one can see that the operation of
\textit{Step (e) Masking} is a bijective
$g_i^3: \{0, 1, 2, 3\} \to \{0, 1, 2, 3\}$.
\end{inparaenum}
According to the above three, one can obtain the conclusion that
the encryption operations of \textit{Step (c)} to \textit{Step (e)} are
bijection $f_i = g_i^3\circ g_i^2 \circ g_i^1$, where
$f_i: \{A, G, C, T\}\to \{0, 1, 2, 3\}$.
Note that, the three bijective $g_i^1, g_i^2$ and $g_i^3$ are satisfy
the Watson-Crick complement rule. Therefore, $f_i$ is equivalent to
a DNA map rule. Thus, this property is proved.
\end{proof}

\begin{Property}
\label{pro:ac}
If $N^g_i = N^b_i$, one has $D^b_i = C$.
\end{Property}
\begin{proof}
The \textit{Step (b) Addition} shows that $D^b_i = N^b_i - N^g_i$.
From the substraction operation of
Table~\ref{tab:AdditionAndSubtraction}, one has $D^b_i = C$.
Thus, this property is proved.
\end{proof}

From Property \ref{pro:bijective}, one knows that a sequence of
DNA map rule $\{h_i\}_{i=1}^{4L}$ is the equivalent secret
key about $k_2$, $\{z_i\}_{i=1}^{4L}$ and $\{t_i\}_{i=1}^{4L}$.
Then, the seemingly complex encryption procedures of \textit{Step (c)}
to \textit{Step (e)} are equivalent to the following simple
decoding step:
\begin{itemize}
\item \textit{Decoding}. For $i=1\sim 4L$, carry out the DNA map rule
$h_i$ to decode $N^r_i, N^g_i$ and $N^b_i$ as $r'_i, g'_i$
and $b'_i$, respectively.
\end{itemize}
Obviously, $N^r_i, N^g_i$ and $N^b_i$ are decoded with the same DNA map
rule $h_i$. Making use of Property \ref{pro:ac}, we only need to
find a element $(r'_{i}, g'_{i}, b'_{i})$, which satisfies
$g'_{i} = b'_{i}$, then one can derive the result that $b_{i}$ maps to $C$.
We denote the relationship between $C$ and $b_{i}$ as
$\Map(C) = b_{i}$. As $C$ and $G$ are complementary, $\Map(G) = (3-\Map(C))$.
Then, one can obtain the scope of $k_1$,
\begin{linenomath}
\begin{equation}
\nonumber
k_1 \in
\begin{cases}
\{3, 4\}, &\mbox{if } \Map(C) = 0,\\
\{1, 7\}, &\mbox{if } \Map(C) = 1,\\
\{2, 8\}, &\mbox{if } \Map(C) = 2,\\
\{5, 6\}, &\mbox{if } \Map(C) = 3,\\
\end{cases}
\end{equation}
\end{linenomath}
and
\begin{linenomath}
\begin{equation}
\nonumber
\{\Map(A), \Map(T)\} = \{0, 1, 2, 3\} \setminus \{\Map(C), \Map(G)\}.
\end{equation}
\end{linenomath}
Table~\ref{tab:at} shows the values of $(D^r_i, D^g_i, D^b_i)$
and its corresponding $(N^r_i, N^g_i, N^b_i)$ can distinguish
$\Map(A)$ from $\Map(T)$. For example, assume that one has
$(r_j, g_j, b_j) = (\Map(C), x, x)$, where $x \in \{\Map(A), \Map(T)\}$
and its corresponding $(D^r_j, D^g_j, D^b_j)$ is $(C, A, A)$ or $(C, T, T)$.
If we find that $r'_j = b'_j$, then one has $\Map(T) = x$, otherwise $\Map(A) = x$.
By observing the Table~\ref{tab:at}, we can know that
when $(D^r_j, D^g_j, D^b_j)$ have
the form shown in Table~\ref{tab:at}, $r'_j = b'_j$ is the
important condition to distinguish $\Map(A)$ from $\Map(T)$.

\begin{table*}[htbp]
\centering
\caption{The values of $(D^r_i, D^g_i, D^b_i)$ and its corresponding
$(N^r_i, N^g_i, N^b_i)$ can distinguish $\Map(A)$ from $\Map(T)$.}
{\begin{tabu}{|c|c|[1pt]c|c|[1pt]c|c|}
\hline
$(D^r_i, D^g_i, D^b_i)$ & $(N^r_i, N^g_i, N^b_i)$ &
$(D^r_i, D^g_i, D^b_i)$ & $(N^r_i, N^g_i, N^b_i)$ &
$(D^r_i, D^g_i, D^b_i)$ & $(N^r_i, N^g_i, N^b_i)$ \\\tabucline[1pt]{-}
$(C, A, A)$ & $(A, T, G)$ & $(T, G, A)$ & $(A, C, A)$ & $(A, A, G)$ & $(T, C, G)$\\\hline
$(C, T, T)$ & $(T, C, T)$ & $(A, G, T)$ & $(C, A, G)$ & $(T, T, G)$ & $(C, A, C)$\\\tabucline[1pt]{-}
$(C, A, T)$ & $(A, G, A)$ & $(A, C, T)$ & $(A, T, C)$ & $(A, T, G)$ & $(G, A, C)$\\\hline
$(C, T, A)$ & $(T, G, C)$ & $(T, C, A)$ & $(T, A, T)$ & $(T, A, G)$ & $(G, C, G)$\\\tabucline[1pt]{-}
$(C, C, A)$ & $(C, A, T)$ & $(A, G, G)$ & $(C, T, A)$ & $(A, A, T)$ & $(T, G, A)$ \\\hline
$(C, C, T)$ & $(C, T, C)$ & $(T, G, G)$ & $(A, T, A)$ & $(T, T, A)$ & $(C, G, C)$ \\\tabucline[1pt]{-}
$(C, G, A)$ & $(G, C, A)$ & $(A, C, G)$ & $(A, G, T)$ & $(A, T, T)$ & $(G, C, T)$ \\\hline
$(C, G, T)$ & $(G, A, G)$ & $(T, C, G)$ & $(T, G, T)$ & $(T, A, A)$ & $(G, T, G)$ \\\hline
\end{tabu}}
\label{tab:at}
\end{table*}

Once $k_1$ has been confirmed, one can obtain
$\{h_i\}_{i=1}^{4L}$. The above analysis shows that
$h_i$ can be determined if and only if at least one of the three sets,
$\{N^r_i, N^g_i\}$, $\{N^r_i, N^b_i\}$ and $\{N^g_i, N^b_i\}$,
in the set $\{\{A, G\}, \{A, C\}, \{T, G\}, \{T, C\}\}$.
One has the set $\mathbf{R}$, where
\begin{linenomath}
\begin{equation}
\nonumber
\begin{split}
\mathbf{R} = \{
   &(T, A, A), (G, G, C), (C, G, G), (A, A, T), \\
   &(G, C, G), (C, G, C), (A, T, A), (T, A, T), \\
   &(C, C, G), (A, T, T), (T, T, A), (G, C, C), \\
   &(A, A, A), (T, T, T), (G, G, G), (C, C, C)\}.
\end{split}
\end{equation}
\end{linenomath}
Therefore, if $(N^r_i, N^g_i, N^b_i) \in \mathbf{R}$, $h_i$
cannot be determined. To help determine
$\{h_i\}_{i=1}^{4L}$ completely, we make use of the following property.
\begin{Property}
\label{pro:class}
Given the value of $k_2$, the scope of $h_i$ can be narrowed via
\begin{linenomath}
\begin{equation}
\label{eq:property}
h_i \in
\begin{cases}
\{1, 3, 6, 8\}, &\mbox{if } k_2 \in \{1, 3, 6, 8\},\\
\{2, 4, 5, 7\}, &\mbox{if } k_2 \in \{2, 4, 5, 7\}.\\
\end{cases}
\end{equation}
\end{linenomath}
\end{Property}
\begin{proof}
It is easy to obtain Eq.~(\ref{eq:property}) from
Table~\ref{tab:property}, which lists the values of $h_i$
with all possible different values of $z_i, k_2$ and $t_i$.
The proof is thus completed.
\end{proof}

\begin{table*}[htbp]
\centering
\caption{The value of $h_i$ corresponding to the values of $z_i, k_2$ and $t_i$.}
{\begin{tabular}{*{9}{|c}|}
\hline
\multirow{2}{*}{$k_2$} &\multicolumn{4}{c|}{$z_i = 0$}
&\multicolumn{4}{c|}{$z_i = 1$} \\
\cline{2-9}
&$t_i=0$ &$t_i=1$ &$t_i=2$ &$t_i=3$ &$t_i=0$ &$t_i=1$ &$t_i=2$ &$t_i=3$\\\hline
$1$ &$1$ &$3$ &$6$ &$8$ &$8$ &$6$ &$3$ &$1$ \\\hline
$2$ &$2$ &$5$ &$4$ &$7$ &$7$ &$4$ &$5$ &$2$ \\\hline
$3$ &$3$ &$1$ &$8$ &$6$ &$6$ &$8$ &$1$ &$3$ \\\hline
$4$ &$4$ &$7$ &$2$ &$5$ &$5$ &$2$ &$7$ &$4$ \\\hline
$5$ &$5$ &$2$ &$7$ &$4$ &$4$ &$7$ &$2$ &$5$ \\\hline
$6$ &$6$ &$8$ &$1$ &$3$ &$3$ &$1$ &$8$ &$6$ \\\hline
$7$ &$7$ &$4$ &$5$ &$2$ &$2$ &$5$ &$4$ &$7$ \\\hline
$8$ &$8$ &$6$ &$3$ &$1$ &$1$ &$3$ &$6$ &$8$ \\\hline
\end{tabular}}
\label{tab:property}
\end{table*}

Referring to Property \ref{pro:class}, one can see that
if we know the scope of $k_2$ and one of the $\Map(A), \Map(G), \Map(T)$
and $\Map(C)$, $h_i$ can be determined
via checking Table~\ref{tab:EncodingRules}.

Assume that a plain-image $\mathbf{I}= \{I_i\}_{i=1}^{L}$
and the corresponding cipher-image $\mathbf{I'}= \{I'_i\}_{i=1}^{L}$
are available, and then we can obtain $\mathbf{I}_b = \{(r_i, g_i, b_i)\}_{i=1}^{4L}$
and $\mathbf{I}'_b = \{(r'_i, g'_i, b'_i)\}_{i=1}^{4L}$.
The detailed procedure of recover $k_1$ and $\{h_i\}_{i=1}^{4L}$
can be described as follows.
\begin{itemize}[noitemsep, nolistsep]
\item \textbf{Step 1}: Search for a element in $\mathbf{I}'_b$ whose
value satisfies $g'_{i_0} = b'_{i_0}$, and then obtain
$\Map(C) = b_{i_0}$, $\Map(G) = 3 - b_{i_0}$.

\item \textbf{Step 2}: Search for a element in $\mathbf{I}_b$ whose
corresponding $(D^r_{i_1}, D^g_{i_1}, D^b_{i_1})$ has the form
as the Table~\ref{tab:at} shown, and then obtain $\Map(A)$.
Thus, the value of $k_1$ is recovered and we can further obtain the
DNA sequence $\{(N^r_{i}, N^g_{i}, N^b_{i})\}_{i=1}^{4L}$.

\item \textbf{Step 3}: Search for a element in
$\{(N^r_{i}, N^g_{i}, N^b_{i})\}_{i=1}^{4L}$ whose value
satisfies at least one of the three sets $\{N^r_{i_2}, N^g_{i_2}\}$,
$\{N^r_{i_2}, N^b_{i_2}\}$ and $\{N^g_{i_2}, N^b_{i_2}\}$ in the set
$\{\{A, G\}, \{A, C\}, \{T, G\}, \{T, C\}\}$, and then obtain the scope of $k_2$.

\item \textbf{Step 4}: For $i = 1 \sim 4L$, get $h_i$ according to the scope of
$k_2$, $N^r_{i}$ and $r'_i$.

\end{itemize}

Now, let's analyze the performance of the above attack.
We know that $(r_i, g_i, b_i)$ has only $4^3 = 64$ different values,
among them, there are $4\times 4 = 16$ kinds of $(r_i, g_i, b_i)$ can
be used to determine $\Map(C)$,
and $24$ kinds of $(r_i, g_i, b_i)$ can be used to derive $\Map(A)$,
otherwise,
the scope of $k_2$ can be determined by
$64 - |\mathbf{R}| = 48$ kinds of $(r_i, g_i, b_i)$.
As the plaintext is chosen from natural images, the value of pixels
follows Gaussian distribution. Thus, one can assure that
the value of $k_1$ and the scope of $k_2$ can be determined with an very extremely high probability.
The computational complexity of the attack is $O(4L)$.

\begin{figure}[!htb]
\centering
\subfigure[]{
    \label{figure:kp:peppers}
    \begin{minipage}[t]{\imagewidth}
    \centering
    \includegraphics[width=\imagewidth]{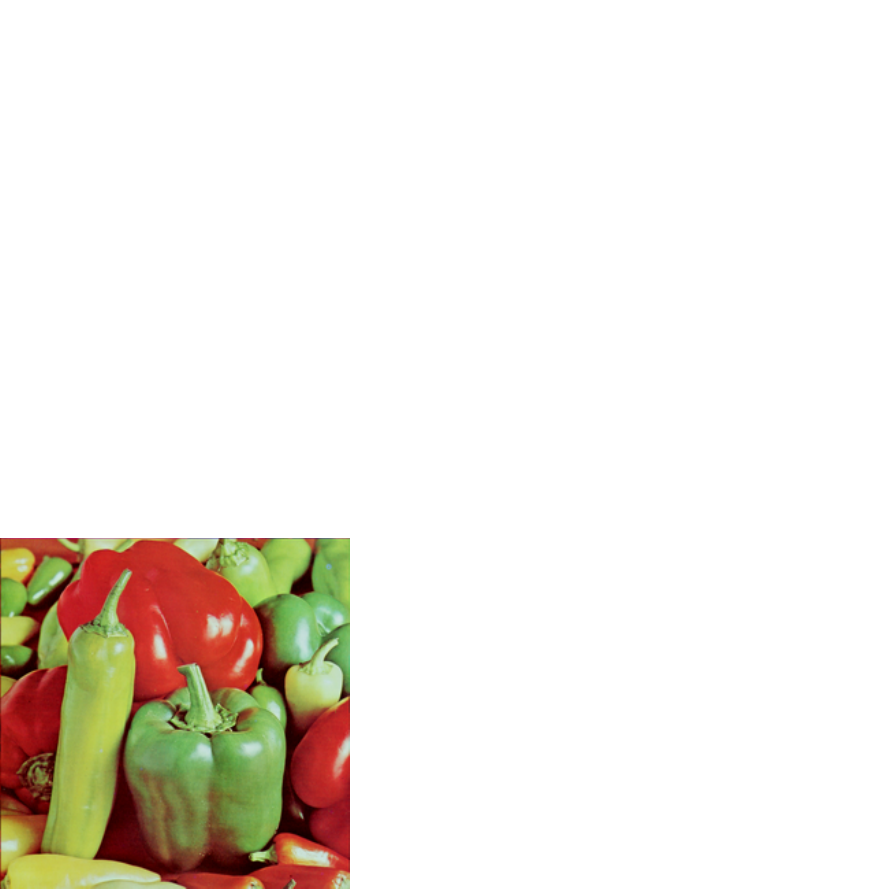}
    \end{minipage}}
\subfigure[]{
    \label{figure:kpa:cpeppers}
    \begin{minipage}[t]{\imagewidth}
    \centering
    \includegraphics[width=\imagewidth]{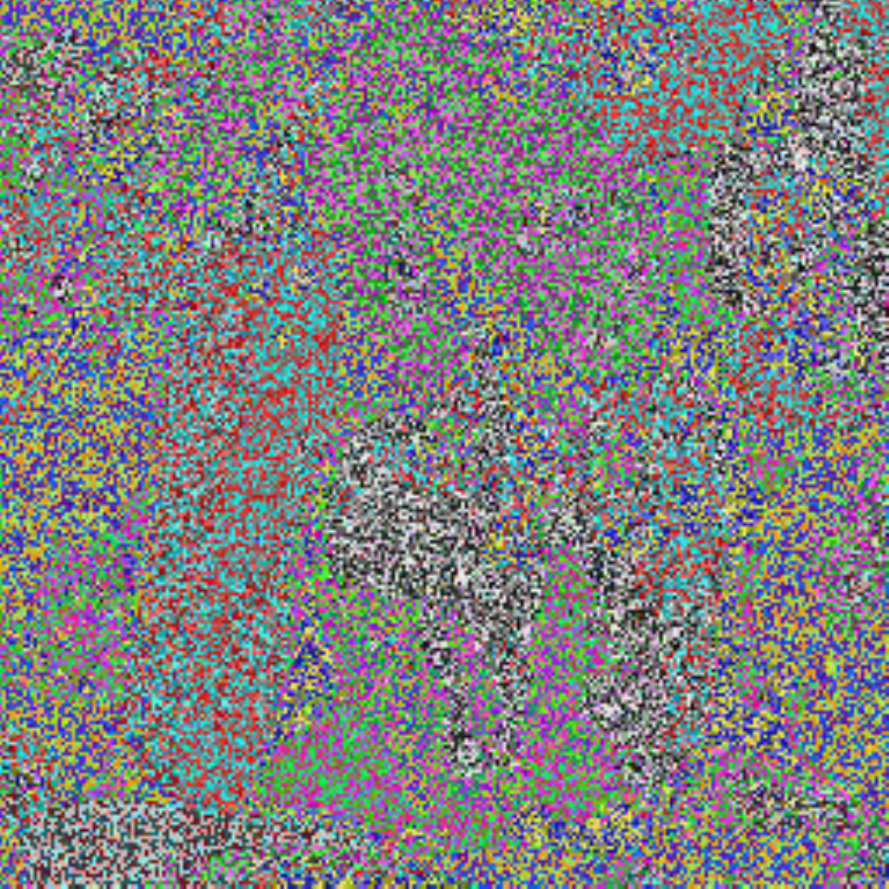}
    \end{minipage}}
\caption{One known plain-image and the corresponding cipher-image:
(a) known plain-image ``Peppers";
(b) cipher-image of Fig.~\ref{figure:kp:peppers}.}
\label{figure:kp}
\end{figure}

\begin{figure}[!htb]
\centering
\subfigure[]{
    \label{figure:kpa:clenna}
    \begin{minipage}[t]{\imagewidth}
    \centering
    \includegraphics[width=\imagewidth]{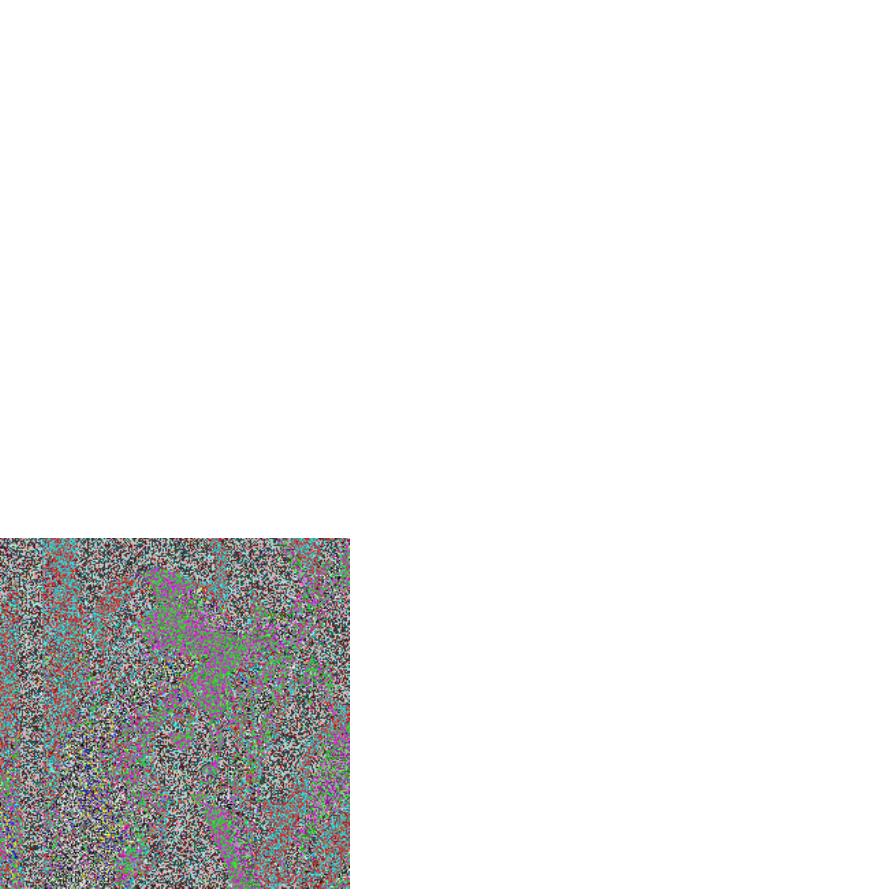}
    \end{minipage}}
\subfigure[]{
    \label{figure:kpa:dlenna}
    \begin{minipage}[t]{\imagewidth}
    \centering
    \includegraphics[width=\imagewidth]{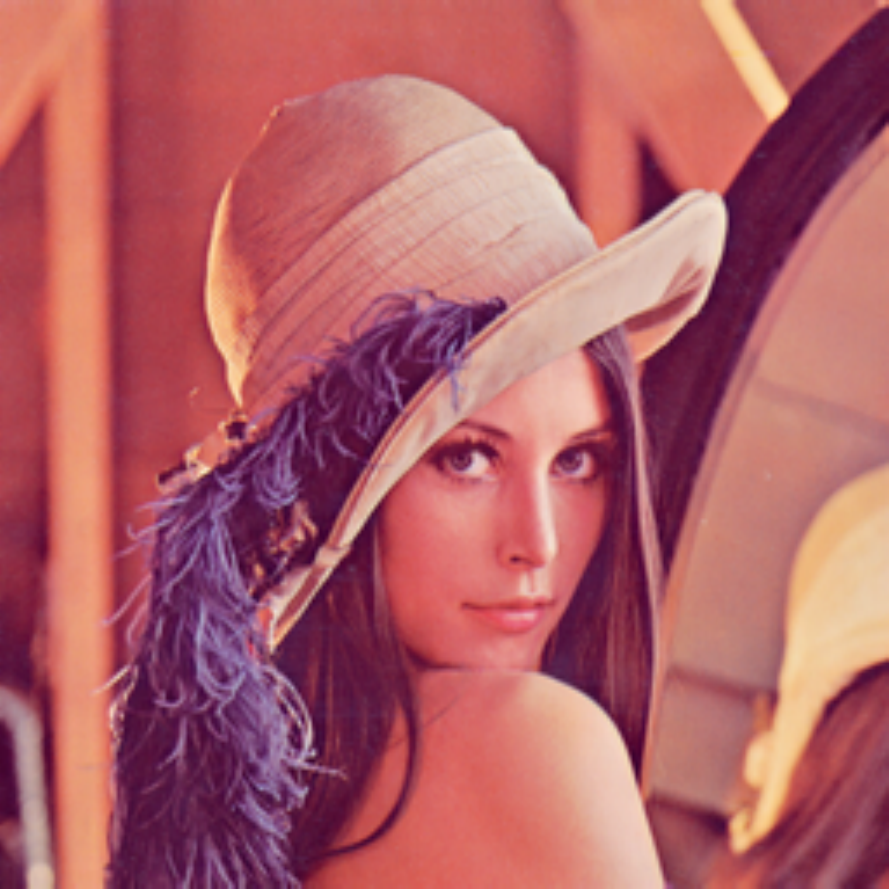}
    \end{minipage}}
\caption{known-plaintext attack:
(a) cipher-image of plain-image ``Lenna";
(b) the recovered image of Fig.~\ref{figure:kpa:clenna}.}
\label{figure:kpa}
\end{figure}

To verify the feasibility of the above known-plaintext attack,
some experiments were performed with plain-images of size
$256\times 256$ (height $\times$ width). The same secret key
used in \cite[Sec.~4]{Liu20121240} was adopted:
$k_1=1$, $k_2=7$, $(x_0, \mu_0)=(0.501, 3.81)$ and
$(x'_0, \mu'_0)=(0.401, 3.68)$.
Figure \ref{figure:kp} shows a plain-image ``Peppers"
and the corresponding cipher-image. We can get
the equivalent secret key $k_1$ and $\{h_i\}_{i=1}^{4L}$.
Finally, the obtained equivalent secret key is used to decrypt another
cipher-image encrypted by the same secret key, as shown in
Fig.~\ref{figure:kpa:clenna}, and the recovery result is shown in
Fig.~\ref{figure:kpa:dlenna}, which is identical with the original plain-image.

\subsection{Two other security defects}

\begin{itemize}
\item \textit{Low Sensitivity with Respect to Changes of Plaintext}

It is suggested in \cite[Sec.~4]{Alvarez2006some} that a new
cryptosystem should be sensitive with respect to plaintext.
But the image encryption algorithm under study is actually very
far from the desired property. As well known in cryptography,
the property is termed as avalanche effect. The desired property
is especially important for secure image encryption algorithms since image
and its watermarked versions, which a slight change of the original image,
are encrypted often at the same time.
This avalanche effect is quantitatively measured by how many
ciphertext bits will change when only one plaintext bit is modified.
As there is no diffusion operation to 
spread the changes out to influence more bits of the different location in
corresponding cipher-image,
the encryption algorithm under study can not reach the desired
state. Obviously, we can easily find that change of a single bit of
plain-image can influence four bits of the corresponding
cipher-image at most.

\item \textit{Low Sensitivity with Respect to Changes of Secret Key}

In \cite[Sec.~5.1.2]{Liu20121240}, the author claimed that the image
encryption algorithm under study has the secret key sensitivity.
However, this claim is questionable as following reasons:
\begin{inparaenum}[(1)]
\item the encryption procedures of \textit{Step (c)} to \textit{Step (e)}
are equivalent to a simple decoding procedure;
\item the confusion procedure \textit{Step (b) Addition} is independent
to secret key;
\item strong redundancy exists among neighboring pixels and the
correlations between $R$, $G$, $B$ components.
\end{inparaenum}
To show this defect clearly, 
a randomly secret key $k_1=2, k_2=5$, $(x_0, \mu_0)=(0.611, 3.781)$
and $(x'_0, \mu'_0)=(0.301, 3.78)$ are used to decrypt the
cipher-image shown in Fig.~\ref{figure:kpa:clenna}, and
the result is shown in Fig.~\ref{figure:error:dlenna}.
Considering that human eyes have a powerful capability of
correcting errors and recognizing significant information.
It is found that some visual information contained in Fig.~\ref{figure:error:dlenna},
although none of pixels are correct in value.

\begin{figure}[htb]
\centering
\subfigure[]{
    \label{figure:error:dlenna}
    \begin{minipage}[t]{\imagewidth}
    \centering
    \includegraphics[width=\imagewidth]{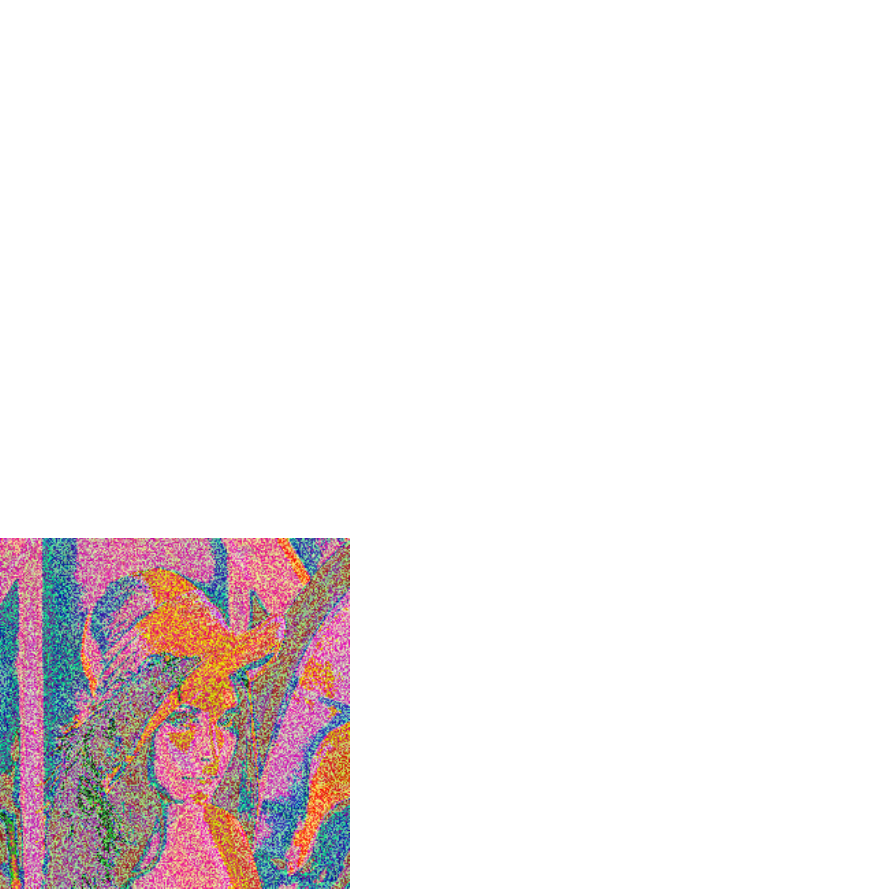}
    \end{minipage}}
\subfigure[]{
    \label{figure:error:dlennar}
    \begin{minipage}[t]{\imagewidth}
    \centering
    \includegraphics[width=\imagewidth]{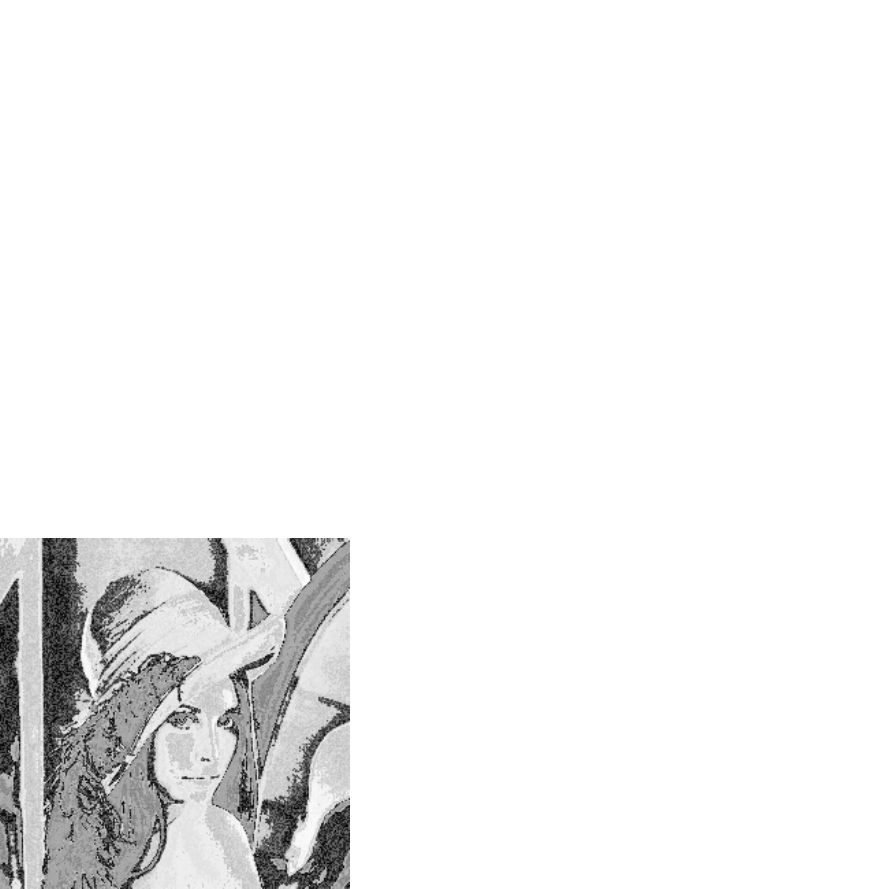}
    \end{minipage}}
\subfigure[]{
    \label{figure:error:dlennag}
    \begin{minipage}[t]{\imagewidth}
    \centering
    \includegraphics[width=\imagewidth]{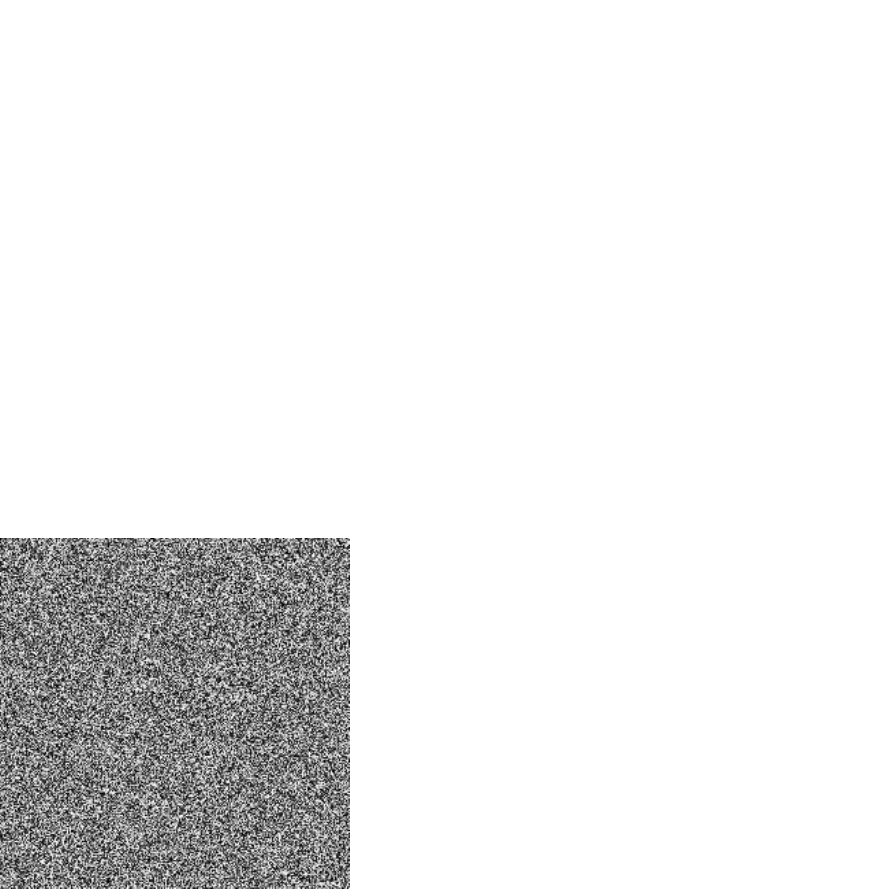}
    \end{minipage}}
\subfigure[]{
    \label{figure:error:dlennab}
    \begin{minipage}[t]{\imagewidth}
    \centering
    \includegraphics[width=\imagewidth]{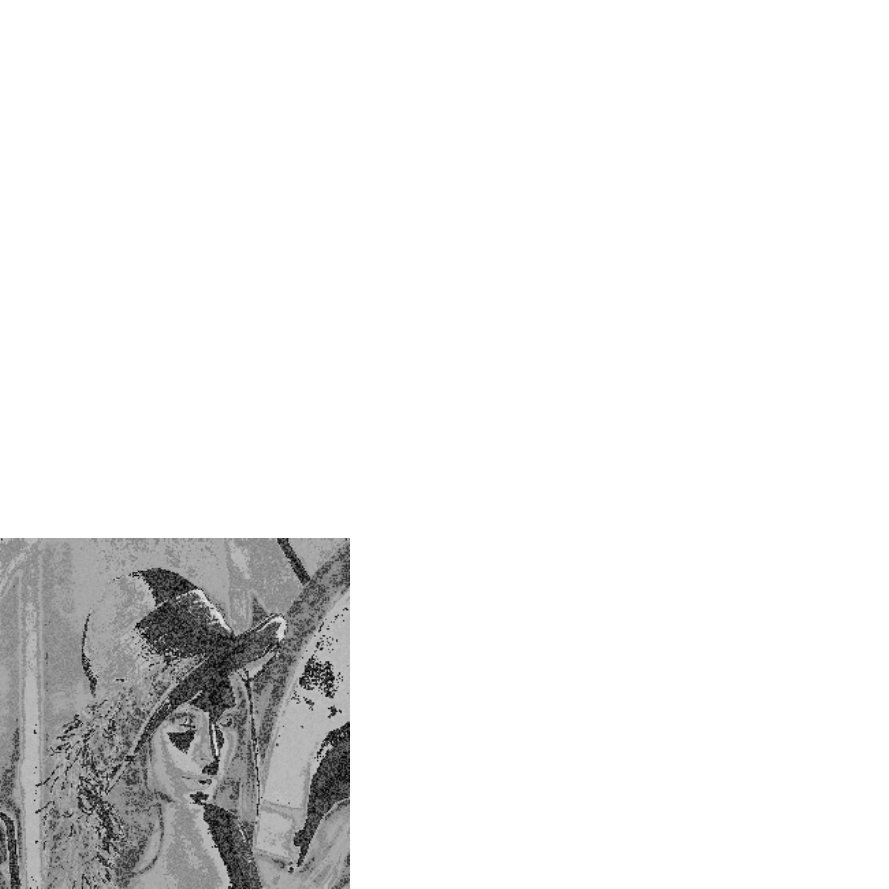}
    \end{minipage}}
\caption{Key sensitivity test:
(a) the error key recovered plain-image from the image shown in Fig.~\ref{figure:kpa:clenna};
(b) R component of the recovered image;
(c) G component of the recovered image;
(d) B component of the recovered image.}
\label{figure:error}
\end{figure}

\end{itemize}

\section{Conclusion}
This paper re-evaluated the security of a RGB image encryption algorithm
based on DNA encoding and chaos map proposed in \cite{Liu20121240}.
It was found that the seemingly complex encryption algorithm can be
effectively broken with only one known plain-image.
Detailed cryptanalytic investigations are given and some experiments are
made to verify the feasibility of the proposed known-plaintext attack.
In addition, some other security
weaknesses of the encryption algorithm was also shown.
Therefore, we suggest not using it in applications that requires
a high level of security.

\bibliographystyle{elsarticle-num}
\bibliography{OLTBIB}
\end{document}